\documentclass[prl,twocolumn,floatfix,superscriptaddress]{revtex4}
\usepackage{graphicx,citesort,psfrag,amsmath,amssymb}

\begin{document}

\bibliographystyle{apsrev}

\title{Comment on ``Relevant Length Scale of Barchan Dunes''}

\author{Klaus Kroy} \affiliation{Hahn-Meitner Insititut, Glienicker
Str.~100, 14109 Berlin, Germany} \author{Xiang Guo}
\affiliation{Department of Building and Construction, City University
of HongKong, HongKong}

\noindent \textbf{Comment on ``Relevant Length Scale of Barchan Dunes''}

In a recent experimental breakthrough, Hersen \emph{et al.}
\cite{hersen-douady-andreotti:2002} demonstrated that by changing the
agitating medium from air to water, one can obtain, on laboratory
scale, dunes that are downsized copies of desert dunes, thereby
overcoming a major obstacle for their systematic study.  Here we argue
in two steps (\emph{i}), (\emph{ii}) that an alternative data analysis
leads to some conclusions that are qualitatively and quantitatively
different from Hersen \emph{et al.}'s but justify their similarity
hypothesis on different grounds.

It has been a long--standing debate whether the various shapes of
desert dunes share the scale--invariance of the turbulent wind field
that creates them. Careful field measurements of barchan dunes have
recently provided compelling evidence that scale invariance is borne
out only \emph{transverse} to the wind direction, whereas it is
systematically broken \emph{along} the wind direction
\cite{sauermann-etal:2000}. That the responsible characteristic scale
is the windward distance $\ell_s$ over which the sand flux adapts to a
sudden change in wind speed or sand supply, was already vaguely
anticipated by Bagnold. But only recent theoretical developments
succceeded in making this notion fully quantitative and established
its crucial role for systematic shape variations and shape transitions
of aeolian dunes \cite{sauermann-kroy-herrmann:2001}.

(\emph{i}) The analysis of similarities of aeolian and submarine dunes
in Ref.~\cite{hersen-douady-andreotti:2002} suffers from mixing
transverse and longitudinal dune dimensions. From what was said above,
a genuine test of similarity should address the longitudinal geometry
\emph{directly}, thereby avoiding influences
\cite{sauermann-etal:2000} from uncontrolled parameters affecting the
transverse direction.  Recombining the data from
Ref.~\cite{hersen-douady-andreotti:2002}, we can infer the (affine)
aeolian and submarine height--length ($H$--$L$) relations that suggest
a considerably larger scale factor than estimated in
\cite{hersen-douady-andreotti:2002}. Our best estimate for the latter
is however obtained from analysing data
\cite{hersen-douady-andreotti:2002,finkel:59} for the dune migration
speed $V$ as follows.

\begin{figure}[t]
\psfrag{V}[cc]{speed $V$ [mm/min]}
\psfrag{H}[cc]{height $H$ [mm]}
\psfrag{l1}{$L^{-1}$}
\psfrag{l2}{$H^{-1}$}
\psfrag{l3}{submarine}
\psfrag{l4}{rescaled aeolian}
\includegraphics{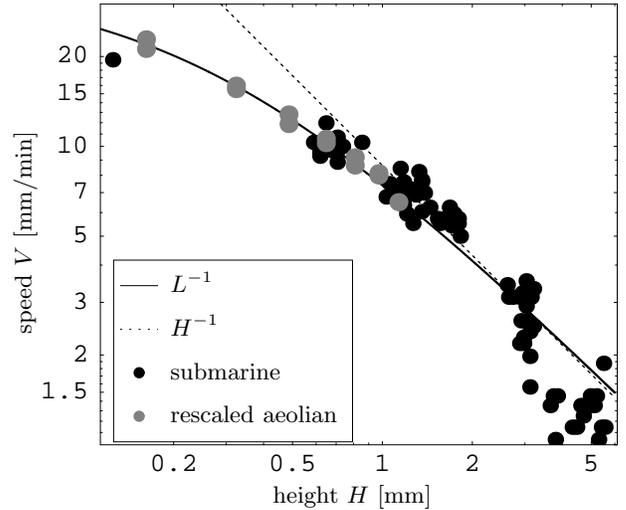}
\caption{Submarine (black \cite{hersen-douady-andreotti:2002}) and
aeolian (gray \cite{finkel:59}; rescaled) migration speed data
compared to $H^{-1}$ (dotted \cite{hersen-douady-andreotti:2002})
failing for small dunes, and $L^{-1}$ (solid
\cite{sauermann-kroy-herrmann:2001}) compatible with the data.}
\label{fig:dunevel}
\end{figure}

(\emph{ii}) The (trivial) fact that for a barchan dune, which looses
practically no sand over its crest, $V$ is proportional to the flux
$q_c$ over the crest divided by its height, is the origin of the
popular relation $V\propto H^{-1}$ adopted in
Ref.~\cite{hersen-douady-andreotti:2002}. Note however that the latter
would only hold for size--independent $q_c$, i.e.\ in the case of
scale--invariant dune shape, inconsistent with the observation of
broken scale invariance crucially underlying the similarity hypothesis
by Hersen \emph{et al}. Yet, one can show on quite general grounds
\cite{sauermann-kroy-herrmann:2001} that (essentially) $V\propto
L^{-1}$, $L$ being the total length of either smooth heaps or barchan
dunes.  Fig.~\ref{fig:dunevel} demonstrates that this (but \emph{not}
$V\propto H^{-1}$) is compatible with \emph{both} aeolian and
submarine data.  To superimpose the data, lengths had to be rescaled
by $6.2\cdot 10^3$ (in line with our recombined $H-L$ relations)
instead of the $1.7\cdot 10^3$ suggested in
\cite{hersen-douady-andreotti:2002}, and times by $2.3 \cdot
10^6$. Assuming $\bar \ell_s^{\rm air}\approx 0.8$~m
\cite{sauermann-kroy-herrmann:2001} for the flux--weighted average of
$\ell_s$ \footnote{While the average saltation length, the scale
$l_{\rm drag}$ introduced in Ref.~\cite{hersen-douady-andreotti:2002},
and $\bar \ell_s$ are all proportional to each other on dimensional
grounds, their absolute values are generally stronlgy diverse due to
distinctly different dependencies on the wind speed
\cite{sauermann-kroy-herrmann:2001}, which prohibits their
identification even for order--of--magnitude estimates.} this implies
$\bar \ell_s^{\rm water}\approx 0.13$\,mm equal to the grain
diameter, as naturally expected for direct fluid entrainment.
In conclusion, we agree with Ref.~\cite{hersen-douady-andreotti:2002}
that there is an intriguing similarity between submarine and aeolian
dunes, but on somewhat different grounds. Further quantitative
confirmation of this similarity by data for small submarine dunes,
backing the highest submarine data point in Fig.~\ref{fig:dunevel},
would be desirable.

\vfill

\noindent Klaus Kroy$^1$ and Xiang Guo$^2$

$^1${\small Hahn-Meitner Institut 

\mbox{}$\,$ Glienicker Str.~100, 14109 Berlin, Germany}

 $^2${\small Department of Building and Construction,  

\mbox{}$\,$  City University of HongKong, HongKong}


\begin{thebibliography}{4}
\expandafter\ifx\csname natexlab\endcsname\relax\def\natexlab#1{#1}\fi
\expandafter\ifx\csname bibnamefont\endcsname\relax
  \def\bibnamefont#1{#1}\fi
\expandafter\ifx\csname bibfnamefont\endcsname\relax
  \def\bibfnamefont#1{#1}\fi
\expandafter\ifx\csname citenamefont\endcsname\relax
  \def\citenamefont#1{#1}\fi
\expandafter\ifx\csname url\endcsname\relax
  \def\url#1{\texttt{#1}}\fi
\expandafter\ifx\csname urlprefix\endcsname\relax\def\urlprefix{URL }\fi
\providecommand{\bibinfo}[2]{#2}
\providecommand{\eprint}[2][]{\url{#2}}

\bibitem[{\citenamefont{Hersen et~al.}(2002)\citenamefont{Hersen, Douady, and
  Andreotti}}]{hersen-douady-andreotti:2002}
\bibinfo{author}{\bibfnamefont{P.}~\bibnamefont{Hersen}},
  \bibinfo{author}{\bibfnamefont{S.}~\bibnamefont{Douady}}, \bibnamefont{and}
  \bibinfo{author}{\bibfnamefont{B.}~\bibnamefont{Andreotti}},
  \bibinfo{journal}{Phys.~Rev.~Lett.} \textbf{\bibinfo{volume}{89}},
  \bibinfo{pages}{264301} (\bibinfo{year}{2002}).

\bibitem[{\citenamefont{Sauermann et~al.}(2000)\citenamefont{S
auermann, Rognon, Poliakov, and Herrmann}}]{sauermann-etal:2000}
\bibinfo{author}{\bibfnamefont{G.}~\bibnamefont{Sauermann}},
\bibinfo{author}{\bibfnamefont{P.}~\bibnamefont{Rognon}},
\bibinfo{author}{\bibfnamefont{A.}~\bibnamefont{Poliakov}},
\bibnamefont{and} \bibinfo{author}{\bibfnamefont{H.~J.}
\bibnamefont{Herr-mann} }, \bibinfo{journal}{Geomorphology}
\textbf{\bibinfo{volume}{36 }}, \bibinfo{pages}{47}
(\bibinfo{year}{2000}).


\bibitem[{\citenamefont{Sauermann et~al.}(2001)\citenamefont{Sauermann, Kroy,
  and Herrmann}}]{sauermann-kroy-herrmann:2001}
\bibinfo{author}{\bibfnamefont{G.}~\bibnamefont{Sauermann}},
  \bibinfo{author}{\bibfnamefont{K.}~\bibnamefont{Kroy}}, \bibnamefont{and}
  \bibinfo{author}{\bibfnamefont{H.~J.} \bibnamefont{Herrmann}},
  \bibinfo{journal}{Phys.~Rev.~E} \textbf{\bibinfo{volume}{64}},
  \bibinfo{pages}{031305} (\bibinfo{year}{2001}); 
  \bibinfo{author}{\bibfnamefont{K.}~\bibnamefont{Kroy}},
  \bibinfo{author}{\bibfnamefont{G.}~\bibnamefont{Sauermann}},
  \bibnamefont{and} \bibinfo{author}{\bibfnamefont{H.~J.}
  \bibnamefont{Herrmann}}, \bibinfo{journal}{Phys.~Rev.~Lett.}
  \textbf{\bibinfo{volume}{88}}, \bibinfo{pages}{054301}
  (\bibinfo{year}{2002}{\natexlab{a}}), and \bibinfo{journal}{Phys.~Rev.~E}
  \textbf{\bibinfo{volume}{66}}, \bibinfo{pages}{031302}
  (\bibinfo{year}{2002}{\natexlab{b}}).

\bibitem[{\citenamefont{Finkel}(1959)\citenamefont{Finkel}}]{finkel:59}
\bibinfo{author}{\bibfnamefont{H.~J.}~\bibnamefont{Finkel}},
 \bibinfo{journal}{J.~Geol.}
\textbf{\bibinfo{volume}{67 }}, \bibinfo{pages}{614}
(\bibinfo{year}{1959}).

\end{thebibliography}
\end{document}